\documentclass[
 aip,
 amsmath,amssymb,
preprint
]{./revtex4-2}
\usepackage{hyperref}
\usepackage{natbib}
\usepackage{graphicx}
\usepackage{dcolumn}
\usepackage{bm}
\usepackage{multirow}
\usepackage{amsmath}
\usepackage{makecell}
\usepackage{mathrsfs}
\usepackage{tikz}

\usepackage[utf8]{inputenc}
\usepackage[T1]{fontenc}
\usepackage{mathptmx}

\begin{document}

\preprint{PRF-submitted/manuscript}

\title{An exhaustive review of studies on bio-inspired convergent-divergent riblets}

\author{Arash Mohammadikarachi}
\affiliation{School of Mechanical Engineering, Pusan National University, 2, Busandaehak-ro 63beon-gil, Geumjeong-gu, Busan, 46241, Republic of Korea}

\author{Mustafa Z. Yousif}
\affiliation{School of Mechanical Engineering, Pusan National University, 2, Busandaehak-ro 63beon-gil, Geumjeong-gu, Busan, 46241, Republic of Korea}

\author{Bagus Nugroho}
\affiliation{Department of Mechanical Engineering, University of Melbourne, Melbourne, 3010 Victoria, Australia}

\author{Hee-Chang Lim}
\email[]{Corresponding author, hclim@pusan.ac.kr}
\thanks{}
\affiliation{School of Mechanical Engineering, Pusan National University, 2, Busandaehak-ro 63beon-gil, Geumjeong-gu, Busan, 46241, Republic of Korea}

\date{\today}

\begin{abstract}
Inspired by the unique textures of shark skin and bird flight feathers and tails, the convergent-divergent surface pattern holds promise in modulating boundary layer structures. This surface pattern exhibits protrusions precisely aligned obliquely (angled in the streamwise direction), often referred to as riblets. These riblets are renowned for their ability to influence the large-scale and very-large-scale structures that dominate the boundary layer. This study seeks to elucidate the influence of convergent-divergent riblets on the boundary layer, with a particular focus on the spanwise direction. We offer a review of research concerning vortex generation physics, emphasizing helicoidal and rotational motions within and adjacent to the riblet valleys. In addition, we examine research, both experimental and numerical, addressing key physical parameters of convergent-divergent riblets, including yaw angle, wavelength, viscous-scaled riblet height, fetch length, and the transition from riblets to a smooth surface. The potential for drag reduction using these bio-inspired riblets is examined. In addition, we delve into the different manufacturing techniques for convergent-divergent riblets. Finally, we discuss the possible commercial applications of the convergent-divergent design.
\end{abstract}

\maketitle


\section{\label{sec:level1}Introduction}

\begin{figure*}
\includegraphics [width=0.8\textwidth]{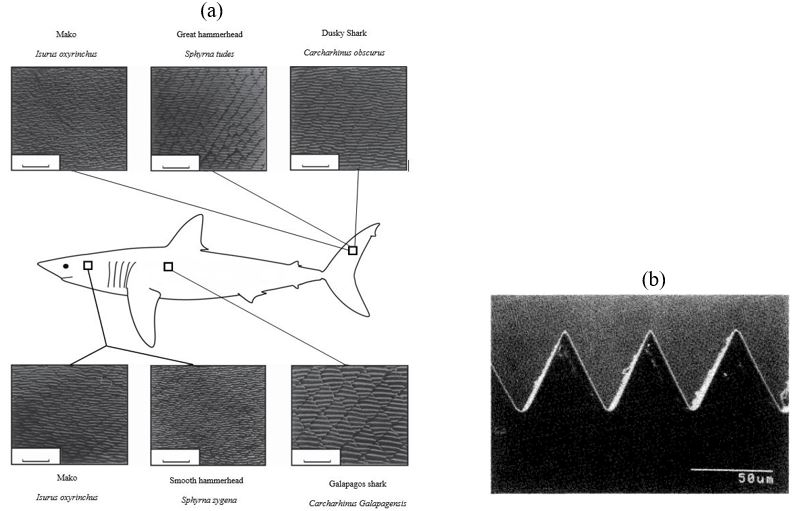}
\caption{(a) Scale view of the skin of fast-swimming sharks (adapted from Reif and Bechert {\it et al.} \cite{reif1985squamation, bechert2000experiments}, (b) cross-section of traditional riblets \cite{walsh1990effect}.} \label{Fig1}
\end{figure*}

As fuel costs continue to rise and global warming becomes an increasingly pressing issue, researchers are driven to explore effective means of reducing energy consumption. Discussions about drag reduction in vehicles and airplanes, which directly impacts energy consumption, have engaged many researchers, engineers, and companies for a long time. Approximately 50\% of the overall drag experienced by civil or commercial transport aircraft can be attributed to skin friction drag \cite{saric2011passive}. The reduction of skin friction drag directly correlates to decreased energy consumption, resulting in lower fuel consumption and reduced emission of greenhouse gases. To overcome all of these challenges, using novel flow control techniques in the new generation of vehicles and aircraft can play a key role in controlling the laminar and turbulent boundary layer flow and drag reduction that significantly decreases energy use. Drag reduction and flow control techniques can be generally classified into two groups: passive techniques that require no energy (vortex generators, splitter plates, riblets, etc.) and active techniques that need energy and a control system (blowing and suction mechanism, jet, and plasma, etc.) \cite{joshi2016review}. These techniques aim to modify the flow around the body, avoid early separation, diminish wake, and encourage mass transfer within the boundary layer. Due to the complexity, high cost of design, and maintenance of the added system, active flow control techniques are less favorable than passive flow control methods \cite{xu2019boundary}. In the last few decades, one nature-inspired passive flow control method in the form of small, groove-like straight structures called riblets has attracted substantial attention \cite{walsh1984optimization, bechert1997experiments}. This unique pattern is inspired by the skin of fast-swimming sharks \cite{reif1985squamation, bechert2000experiments}. It is a promising flow control technique because it can modify wall-bounded turbulent flow and reduce skin friction drag without generating significant form drag \cite{coustols1990synthesis, liu2017reduction}. Figure~\ref{Fig1} shows the groove-like structures on the skin. These tiny structures can dampen and modify cross-flow or the near-wall cycle of streaks and quasi-streamwise vortices and reduce the near-wall velocity gradient \cite{kline1967structure}. Riblets traditionally have a triangular cross-section. However, different cross-sections have emerged recently (i.e. symmetric triangular, asymmetric triangular, blade, trapezoidal, etc.) influencing the turbulent flow structure and drag reduction strength differently \cite{endrikat2021direct, rouhi2022riblet}. Despite the potential of the regular straight riblets to act as a flow control and drag reduction mechanism, their effectiveness only at certain Reynolds numbers constitutes their Achilles heel \cite{bechert1997experiments, garcia2011drag}. 

Beyond this threshold, their drag reduction capability diminishes. Such a situation raises further questions regarding the efficacy of riblets as a flow control and drag reduction technique, because most applied engineering systems that can benefit from riblets, such as aircraft and ships operate at a high Reynolds number range \cite{smits2013wall}. Furthermore, in the last two decades, various reports have shown that at high or very high Reynolds numbers wall-bounded canonical flow (smooth-wall boundary layer, pipe flow and channel flow), the large and very-large-scale motions that reside at the logarithmic region dominate the near-wall cycle of small-scale streaks and streamwise vortices \cite{kim1999very, hutchins2007evidence}. These large-scale and very-large-scale motions are also modulated and strongly influence the near-wall small-scale structure usually controlled by a regular straight riblet. A recent report by Zhang {\it et al.} shows that similar large-scale and small-scale interactions are also observed on the flow over the riblet surface \cite{zhang2020characteristics}. Hence, the riblets will be a more effective flow control mechanism if they can target and affect large and very-large-scale structures. 

\begin{figure*}
\includegraphics[width=0.8\textwidth]{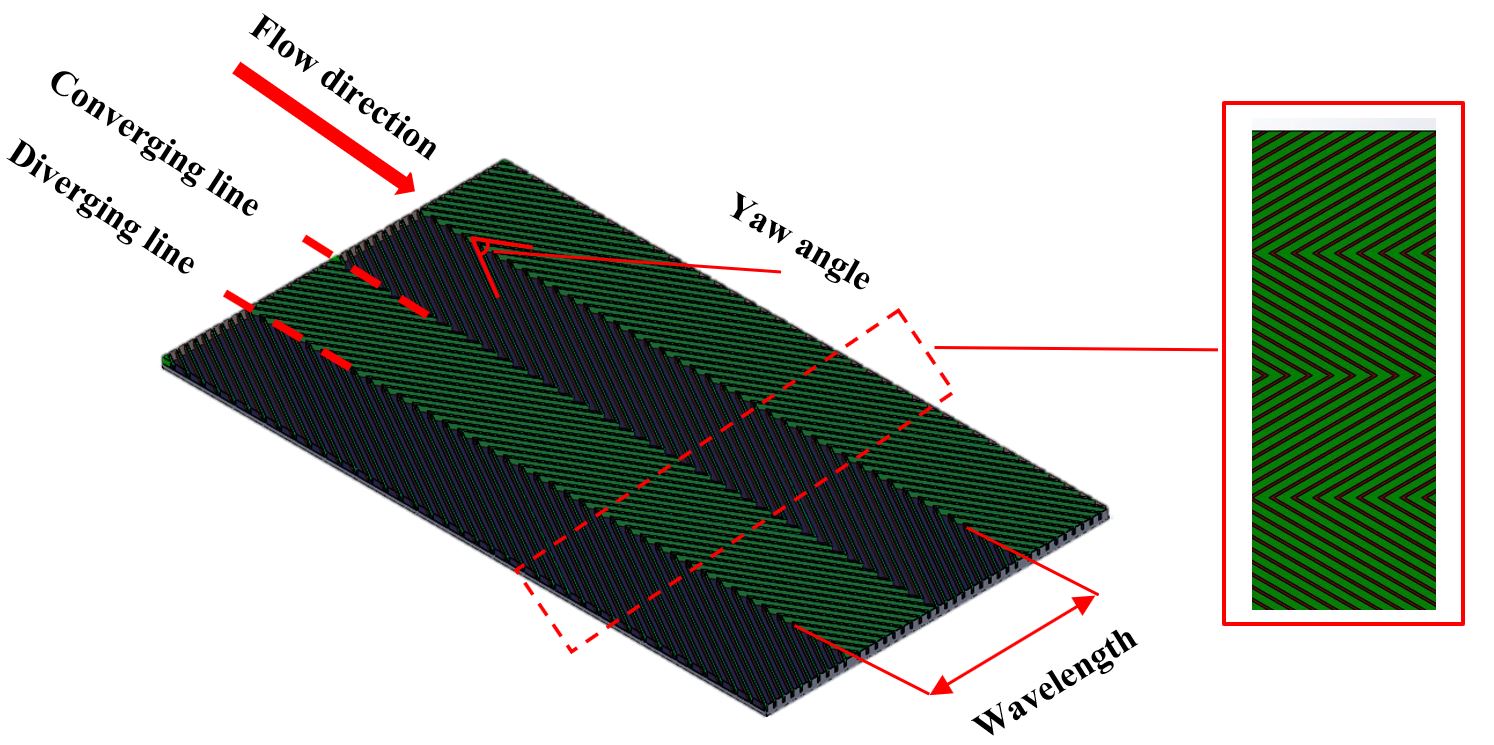}
\caption{Schematic of a surface covered by convergent-divergent riblets.} \label{Fig2}
\end{figure*}

\begin{figure*}
\includegraphics[width=0.6\textwidth]{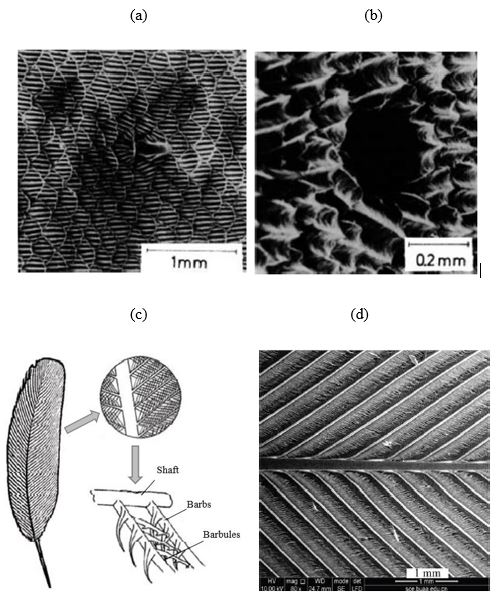}
\caption{(a) Upstream of the sensory receptor of sharks (converging surface), (b) Upstream of the hearing sensors of the sharks (diverging surface) (adapted from Koeltzsch {\it et al.}\cite{koeltzsch2002flow}). (c) Characteristic structure and, (d) Microscopic schematic of bird flight feathers (adapted from Chen {\it et al.} \cite{chen2013biomimetic, chen2014flow}).} \label{Fig3}
\end{figure*}

Koeltzsch {\it et al.} designed a novel class of directional riblets in the form of herringbone or convergent-divergent riblets (referred to as C-D riblets) for first time \cite{koeltzsch2002flow}. These riblets are angled in a streamwise direction called a yaw angle. Figure~\ref{Fig2} illustrates that the combination of spanwise arrangement of the left-tilted and the right-tilted riblets creates C-D riblets. These riblets are inspired by shark skin and bird flight feathers and tails. As seen in Fig.~\ref{Fig3} (a and b), the upstream of shark hearing sensors and shark sensory receptors display diverging and converging patterns respectively \cite{koeltzsch2002flow}, and Fig.~\ref{Fig3} (c and d) illustrates the bird's wing and tail feathers as a diverging pattern \cite{chen2013biomimetic, chen2014flow}. 

Recently, the capability of C-D riblets  to control wall-bounded flow has motivated notable research activities. Using C-D microgrooves is a contemporary and innovative bio-inspired passive flow control method that offers advantages over earlier generations of bio-inspired surfaces. Unlike longitudinal riblets, C-D microgrooves can distinctly impact boundary layer flows, vortical structures, near-wall motions and flow separation \cite{liu2017reduction}. Furthermore, they can target the large and very large-scale features that dominate the boundary layer at high Reynolds numbers \cite{nugroho2013large}. 

\begin{figure*}
\includegraphics[width=0.9\textwidth]{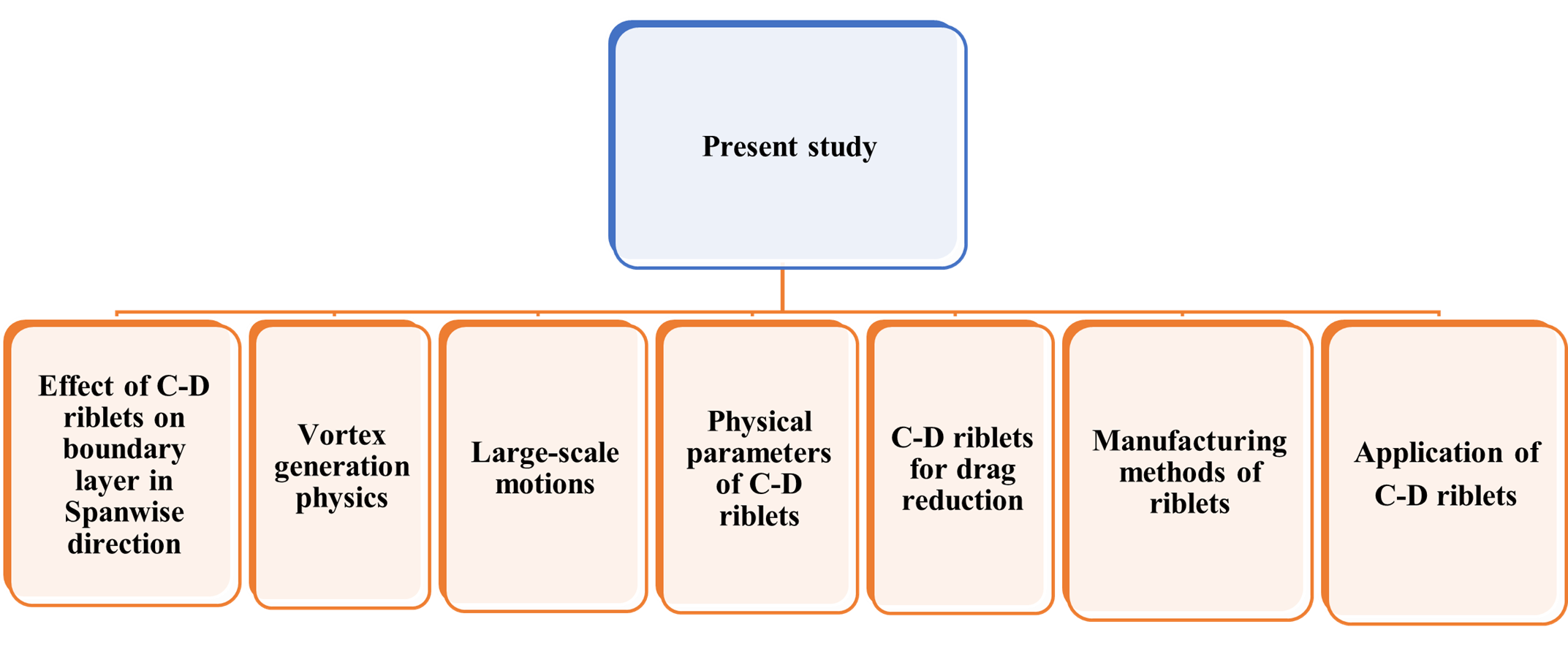}
\caption{An overview of the topics discussed in this study.}\label{Fig4}
\end{figure*}

Due to the C-D riblet's potential as a more robust flow control mechanism than the standard riblets, there has recently been a significant increase in the investigation of this novel pattern. Hence, there is a need for a systematic study that summarises the research findings and progress related to C-D riblets to provide a clear understanding for both researchers and practicing engineers. To address this, the present review delves into the various characteristics of C-D riblets, providing a comprehensive understanding of their influence on multiple aspects within fluid dynamics. Primarily, this study examines their impact on the boundary layer and sheds light on how these micro-structures modify the flow behavior near solid surfaces. In addition, it investigates the interaction of C-D riblets with the large-scale and very-large-scale structures that reside on the logarithmic region of turbulent boundary layer and uncovering their potential to influence the overall flow patterns and turbulence in a given system. This study also encompasses an in-depth analysis of different effective physical parameters that govern the performance of C-D riblets, offering valuable insights into their design and optimisation. Furthermore, the applications where C-D riblets can be deployed for flow separation control is discussed. Moreover, the manufacturing methods of C-D riblets are also thoroughly explained, covering both traditional techniques and emerging advanced processes, enabling researchers and engineers to implement these micro-structures effectively in practical applications. Finally, the possible commercial applications of this surface pattern are expressed. Overall, the present study is a pivotal source of knowledge, paving the way for further advancements and applications of C-D riblets in diverse engineering and commercial industries. Figure~\ref{Fig4} presents an overview of the topics related to C-D riblets discussed in this study.

\section{\label{sec:level1}The general effect of C-D pattern on the spanwise direction of boundary layer (including upwelling-downwelling motions and counter-rotating roll mode)}

\begin{figure*}
\includegraphics[width=0.7\textwidth]{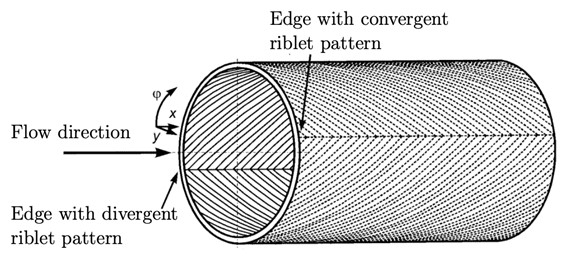}
\caption{convergent and Divergent riblet patterns inside the pipe flow \cite{koeltzsch2002flow}}\label{Fig5}
\end{figure*}

Convergent-divergent riblets can induce alterations in the turbulent boundary layer in the spanwise direction. Koeltzsch {\it et al.} applied C-D pattern riblets inside the surface of a turbulent pipe flow (See Figure~\ref{Fig5}) \cite{koeltzsch2002flow}. The results indicate that this unique pattern can generate large-scale azimuthal variations in the mean velocity and turbulence intensity. To the best of our knowledge, this is the first time that conventional riblets have been modified to have a unique direction. Koeltzsch {\it et al.} discovered that the time-averaged streamwise velocity and the root mean square velocity above the convergent and divergent riblet textures are considerably different near the wall. Clearly, convergent riblet patterns decrease time-averaged velocity and increase velocity fluctuations, whereas divergent riblet patterns exhibit the opposite trend. Nugroho {\it et al.} extended the studies done by Koeltzsch {\it et al.} They designed a plate coated with C-D roughness to check the large-scale spanwise periodicity in a turbulent boundary layer via hot-wire and cross-wire anemometry techniques \cite{nugroho2013large, nugroho2014roll, nugroho2015dissecting, kevin2014wall}. 

\begin{figure}[b]
\includegraphics[width=0.7\textwidth]{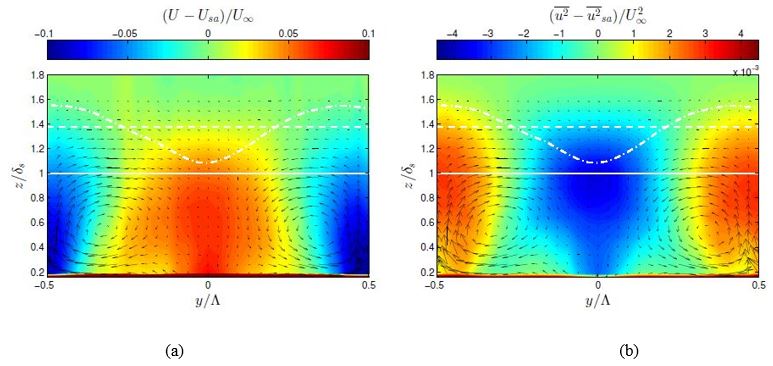}
\caption{Spanwise variation of (a) averaged mean velocity and (b) averaged turbulent intensity \cite{nugroho2013large, nugroho2014roll}. The horizontal solid line, the dot-dashed line, and the dashed line show the boundary layer thickness over the smooth surface, the boundary layer thickness over the C-D riblets, and the spanwise average thickness of the boundary layer respectively. The converging area is located at $y/\Lambda=0.5$,  and -0.5, while the diverging area is at $y/\Lambda=0$} \label{Fig6}
\end{figure}

Different experimental parameters were employed based on various effective physical parameters such as yaw angle, and riblet height. Figure~\ref{Fig6} shows the general effect of C-D riblets on turbulent flow, which illustrates the spanwise variation of mean velocity and turbulence intensity due to C-D riblets. These riblets profoundly modified the mean velocity and turbulence intensity in the spanwise direction. Here over the diverging region, the mean velocity is higher than that of the converging region, whereas the turbulence intensity is lower. This behavior can be explained based on the velocity vectors shown in Fig.~\ref{Fig6}. Over the diverging region, there is a downward flow that directs the high-speed and low-turbulence flow from the upper region of the boundary layer towards the wall. Conversely, the reverse is the case over the converging region, with an upward flow pushing the low-speed and high-turbulent flow from the wall toward the edge of the boundary layer. The combination of this upwelling and downwelling flow creates large-scale counter-rotating vortices. Fig.~\ref{Fig6} also indicates that C-D riblets can significantly change the boundary layer in the spanwise direction. Such variation is remarkable, considering that the riblet height is around 100 times smaller than the boundary layer thickness. Kevin {\it et al.} \cite{kevin2017cross} and Xu {\it et al.} \cite{xu2019statistical} used Particle Image Velocimetry (PIV) measurement to study the vortical structures in the turbulent boundary layer over a flat plate coated with C-D riblets and to provide a clearer visualization of the secondary flow. Their statistical results show a similar behaviour as those by  Nugroho et al \cite{nugroho2013large}, in which the fluid flow over the DL and CL has downwelling and upwelling motions respectively. These in turn cause the local streamwise velocity to increase over the DL, whilst it reduces over the CL.

\section{Vortex generation physics (including helicoidal and rotational motions).}
Different complex flow structures can be produced due to the particular arrangement of the nature-inspired C-D riblets. The profound effect of C-D riblets on vortices, especially near-wall motions, has become an attractive topic for scholars. Because of the distinct fabrication of the riblets which have an angle with the streamwise flow, the formation of the secondary flow inside the gap between the riblets as well as over the riblets differs from the regular straight riblet types. Xu {\it et al.} investigated the vortical structures over a plate covered by triangular riblets in the laminar flow \cite{xu2018vortical}. The micro-scale vortices along and inside the valleys of the riblets and the large-scale vortices across the boundary layer were checked to determine the flow topology. The research demonstrates that the fluid flow has a helicoidal motion inside the valleys. 

\begin{figure}[h!]
\includegraphics[width=0.65\textwidth]{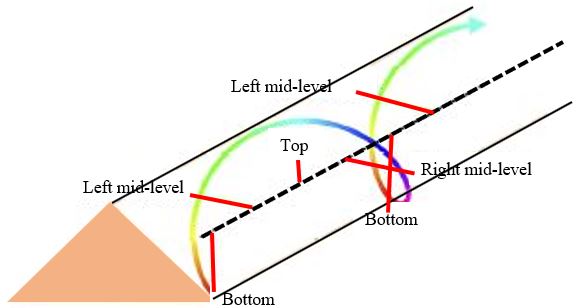}
\caption{The topology of the helical motion along the valleys of the riblets \cite{xu2018vortical}.} \label{Fig7}
\end{figure}

Figure~\ref{Fig7} exhibits that the fluid flow rotates around an axis and travels along the valleys simultaneously. Whenever the fluid moves inside the valleys, it can be assessed as a channel flow under the effect of the cross-flow caused by the riblets' yaw angle. Meanwhile, the axial flow in the valley of the riblets interacts with the cross-flow. As a result of this interaction, the secondary flow and subsequently, helicoidal movement are created.

\begin{figure}[h!]
\includegraphics[width=0.8\textwidth]{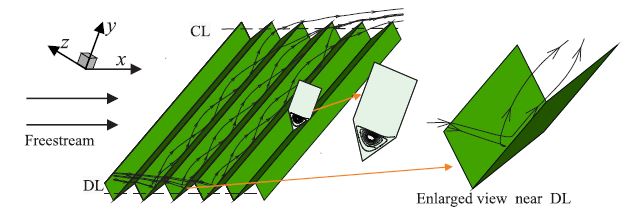}
\caption{Schematic of the streamlines inside the passage of the riblets and in a perpendicular plane \cite{guo2020secondary}. CL and DL are abbreviations of converging and diverging lines respectively.} \label{Fig8}
\end{figure}

Guo {\it et al.} used a plane perpendicular to the valleys of the riblets to illustrate the clockwise helicoidal motion of the streamlines inside the passage of the riblets \cite{guo2020secondary}. This flow motion is generated because of the specific arrangement of the converging and diverging riblets. Notably, another cross-flow with high momentum over the riblets was observed to include two velocity components. One of these velocity components is aligned with the axial component of the helicoidal streamlines inside the valleys. The other component is perpendicular to the riblets and contributes to the rotational component of the helicoidal motion (See Fig.~\ref{Fig8}). 

Xu {\it et al.} investigated the zonal characteristics of uniform streamwise momentum within the boundary layer \cite{xu2019statistical}. They conducted a comparison between the probability distribution function (p.d.f) of the number of uniform momentum zones (UMZs) on both the convergent-divergent riblets and the smooth surface. They calculated the average number of uniform momentum zones ($\bar{N}$umz). A higher value of $\bar{N}$umz over the converging line was observed compared to the smooth surface. Besides, they also found that there is a positive correlation between riblet height and $\bar{N}$umz. For different riblet heights ($h^+$= 8, 14, 12), $\bar{N}$umz=2.567, 2.608, 2.658 were found correspondingly. Beyond the boundary layer, narrower uniform momentum zones were observed on the converging line. An increase in the riblet height could boost this trend.

\section{\label{sec:level1}Large scale motions}
\begin{figure*}
\includegraphics[width=0.5\textwidth]{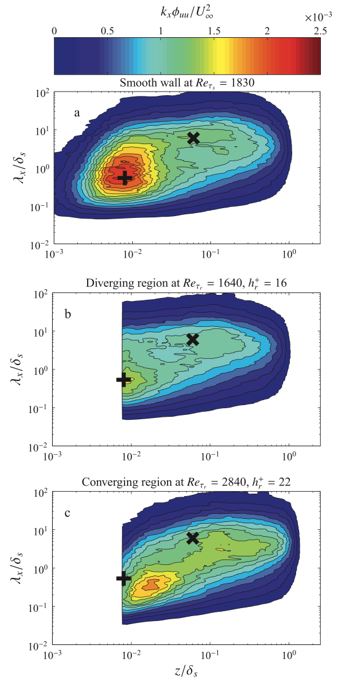}
\caption{Contours of streamwise energy spectra at different skin friction Reynolds and viscous scaled riblet height for (a) smooth, (b) the   diverging region, and (c) the converging region \cite{nugroho2013large}.} \label{Fig9}
\end{figure*}

Further analysis by Nugroho {\it et al} \cite{nugroho2013large} indicates that the large-scale counter-rotating vortices generated by convergent-divergent riblet pattern can significantly influence the large and very-large-scale structures that reside on the logarithmic region of the turbulent boundary layer. The alterations to the turbulent structure are clearly apparent in the pre-multiplied energy spectra plots (See Fig.~\ref{Fig9}) due to the presence of a convergent-divergent surface. These plots are made from individual one-dimensional pre-multiplied energy spectra $K_x$$\Phi_u{}_u$ where $K_x$ is streamwise wavenumber and $\Phi_u{}_u$ is the energy spectra of streamwise velocity fluctuations. Here the horizontal axis is a function of the outer scaled wall-normal position, while the vertical axis is the energetic streamwise length-scale $\lambda_x$/$\delta_s$ in which $\lambda_x$=2$\pi$/$K_x$ and $\lambda_x$ is boundary layer thickness of the smooth wall (see Hutchins and Marusic \cite{hutchins2007evidence}) for further details about the pre-multiplied energy spectra construction)

The smooth wall reference (Figure~\ref{Fig9}a) shows a highly energetic signature near the wall due to the near-wall cycle of streaks and quasi-streamwise vortices, marked by + symbol. This near-wall peak is commonly termed an inner peak. Further from the wall  (around the log region) and at a sufficiently high Reynolds number, it has been observed that there is a much larger scale peak with a length of more than 6$\delta$. In Figure~\ref{Fig9}a, the predicted location is marked by × symbol, and such peak is commonly termed the outer peak. This outer peak is believed to be an energetic imprint of the large (or very large) scale structure. Over the diverging region (See Figure~\ref{Fig9}b), the highly energetic inner peak lies beneath the initial measurement point, hence it is difficult to observe. Further from the wall, there is no noticeable outer peak detected in the diverging spectra. Here the energy spectra's overall magnitude is lesser than that of the smooth wall instance. For the converging region (as depicted in Figure~\ref{Fig9}c), an inner peak is clearly observed, however, it is slightly displaced from the surface compared to the smooth wall scenario. This displacement is likely attributed to the hypothesized secondary flows over the converging that tend to move upwards in this region.

Kevin {\it et al.} used stereoscopic PIV measurement to study the turbulent structures, especially large-scale motions formed over convergent-divergent microgrooves in more detail \cite{kevin2017cross}. For this purpose, they manufactured C-D riblets featuring a trapezoidal cross-section and examined the spanwise periodicity across the riblets. They found that the presence of mean counter-rotating roll modes accompanies this periodicity. They also indicate that significant spanwise variations in Reynolds stress and turbulent kinetic energy with increased quantities are consistently observed above the converging regions. In comparison with the converging and smooth-wall cases, all second-order statistics exhibited lower values over the diverging region. They also studied the streamwise velocity field instantaneously. They indicated the presence of large-scale low-momentum structures above the converging line. These large-scale motions are frequently observed in the spanwise direction, prevalent in other large-scale asymmetric rotational motions. These structures with low momentum on the converging line were observed to be dominant in the flow field. The common flow-down tendency of secondary flows is often associated with the observed occurrence of large-scale free-stream engulfing behavior. They checked the turbulent boundary layer over converging-diverging riblets and illustrated that as a result of the large-scale motions, the free-stream pockets formed above the diverging line in comparison to the smooth wall flow. Bai {\it et al.} used proper orthogonal decomposition (POD) to examine the energetic and large-scale motions of the turbulent boundary layer over the convergent-divergent roughness \cite{bai2021energetic}. They used the dataset of Kevin {\it et al.} \cite{kevin2017cross} corresponding to the instantaneous velocity fields in a cross-stream plane of the turbulent boundary layer at Re=13,000. The mode energy fraction was defined as the contribution of the POD modes and the total turbulent kinetic energy of the turbulent boundary layer. It was observed that whenever the spanwise domain of the POD calculation is set to one wavelength of the converging riblets pattern, the first two modes account for approximately 14.3\% of the total kinetic energy. Alternatively, when the spanwise domain is extended to two wavelengths of the converging riblet pattern, the first four modes demonstrate distinctive characteristics connected to the prominent large-scale low and high-speed structures observed across the converging riblet section. Furthermore, they used the large-magnitude coefficients of the first two POD modes (in the case of one wavelength of converging line) to choose the large-scale low-momentum vortices among the data. Then, they clarified that the primary responsibility for generating mean secondary flow lies with the large-scale low-momentum structures. In summary, the dominant energetic structures in the flow field are associated with the significant coefficients of lower POD modes that exhibit the key characteristics of instantaneous flows. Moreover, combining the dominant large-scale structures in the turbulent boundary layer with the spanwise-heterogeneous roughness yields the mean secondary flows. Kevin {\it et al.} employed particle image velocimetry on different spanwise and wall-normal planes to study the characteristics of large-scale coherent structures in a turbulent boundary layer over C-D riblets \cite{kevin2019turbulent}. They analysed the instantaneous streamwise velocity fields above C-D riblets at multiple time points. They investigated the logarithmic region, during which the long low-momentum structures over the converging line were observed. Although these structures are formed and sustained by the C-D riblets, the characteristics of these turbulent events such as meandering, breaking, and branching are similar to the structures over smooth surfaces. A noteworthy observation in the far outer region is that these low-momentum regions exhibit lateral instability, resulting in pronounced meandering over the yawed riblets, meaning that detached or floating coherence was identified between the upwelling and downwelling regions. The detached or floating coherence originates from the momentum transfer to the outer layer.

\section{\label{sec:level1}The influence of various riblet physical parameters on the flow structure}

Various studies about the converging and diverging riblets in the last decade show that there are several critical parameters that can affect the strength of the spanwise variation, namely: viscous scaled riblet height, riblet yaw angle, wavelength, fetch length, and reversion from C-D riblets to the smooth surface. This section will discuss these physical parameters and illustrate some of the influences.

\subsection{Influence of yaw angle}
\begin{figure*}[h!]
\includegraphics[width=0.7\textwidth]{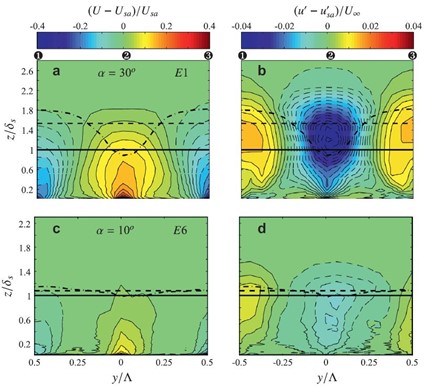}
\caption{The influence of yaw angles 30 (a, b) and 10 (c, d) on spanwise variation for mean velocity (left-side images) and turbulence root-mean-squared fluctuations (right-side images) about the spanwise averaged value for the convergent-divergent riblets \cite{nugroho2013large}.} \label{Fig10}
\end{figure*}

The first important component is the riblet yaw angle, which defines the angle between the riblet and the incoming flow (See Fig.~\ref{Fig10}). Guo {\it et al.} analysed the effect of the yaw angle in the range of 20 to 70 degrees on the boundary layer and secondary flow motion \cite{guo2020secondary}. They checked the streamwise velocity and velocity vectors in the cross-stream plane. Their investigation showed that the strength of the roll motion and the streamwise velocity in the spanwise direction increase with an increase in the yaw angle from 20 to 45 degrees. A higher yaw angle ranging from 45 to 60 degrees reduces the strength of the roll mode and the streamwise velocity in the spanwise direction. Nugroho {\it et al.} investigated the influence of yaw angle on the strength of the secondary flows \cite{nugroho2013large}. They examined the results for yaw angles equal to 10 and 30. As shown in Fig.~\ref{Fig10}, their findings indicated that decreasing the yaw angle reduced the intensity of spanwise variations caused by the surface. When the yaw angle is set to 10, the magnitude of the spanwise variation is notably smaller compared to when it is set to 30. These results indicate that there is a maximum threshold limit in riblet yaw angle. Guo {\it et al.} extended this parameter study and used different yaw angles in the range of 0 to 90 degrees for laminar channel flow to examine the flow structure and separation over a ramp downstream \cite{guo2020control}. At a yaw angle of 0°, C-D riblets transform into longitudinal riblets, and in-plane velocity is not observed above and within the riblet valleys, aligning with the findings of Djenidi et al \cite{djenidi1994laminar}. At 90° yaw angle, C-D riblets transform into transverse riblets, orienting perpendicular to the free-stream flow. The research revealed that due to yaw angle increment, riblets having specific height and spacing exhibit a parabolic relationship between the strength of the secondary flow and the net reduction of the separation zone. The maximum values for both variables occur at a 45-degree yaw angle.

\subsection{Influence of viscous scaled riblet height ($h^+$)}

\begin{figure*}[h!]
\includegraphics[width=0.6\textwidth]{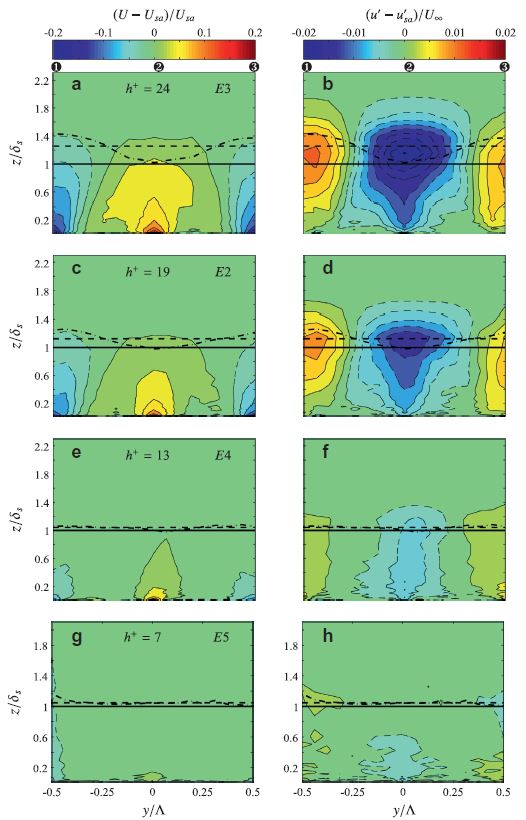}
\caption{Spanwise variation of the mean velocity (left-side images) and turbulence root-mean-squared fluctuations (right-side images) about the spanwise averaged value for the convergent-divergent riblets for (a and b) $h^+$=24, (c and d) $h^+$=19, (e and f) $h^+$=13, (g and h) $h^+$=7 \cite{nugroho2013large}.} \label{Fig11}
\end{figure*}

Another effective parameter related to C-D riblets is the viscous scaled riblet height ($h^+$). There are two methods to obtain different $h^+$, namely: firstly by varying the free stream velocity or downstream location while holding the riblet physical height constant, or secondly by keeping the free stream velocity or downstream location constant and changing the riblets physical height. Nugroho {\it et al.} conducted the first method, by changing the streamwise velocity, to obtain various viscous scaled riblet heights ($h^+$) \cite{nugroho2013large}. Their findings demonstrated how the height of the viscous-scaled riblets influences the strength of the three-dimensionality caused by surface roughness. As $h^+$ decreases to its smallest value, the conditions approach a hydrodynamically smooth case, and the riblets become incapable of inducing significant large-scale three-dimensionality. It can be seen in Fig.~\ref{Fig11}, that with increasing $h^+$, the surface imposes a greater three-dimensional effect on the mean velocity and turbulent intensity. This leads to a more noticeable difference in boundary layer thickness between the diverging and converging regions. 

In the work done by Xu {\it et al.}, the free stream velocity was fixed like the baseline turbulent boundary layer and the physical height (h) was changed to vary ($h^+$) \cite{xu2019statistical}. They conducted the experimental study of turbulent boundary layer over C-D riblets by using PIV measurements. The tests were carried out for three heights ($h^+$=8, 14, and 20) at Re=723. Their results are similar with Nugroho et al \cite{nugroho2013large}, in which an increase in ($h^+$) would result in stronger spanwise variation. 
Beyond looking at the influence of $h^+$, the investigation of Xu et al \cite{xu2018vortical} focused on studying the population of spanwise prograde and retrograde vortices. Their results indicate that when the riblet height is set to 8, the population densities of spanwise vortices moving in the prograde and retrograde directions show an increase of over 50\% in the converging region compared to the smooth-wall scenario. This suggests a substantial rise in turbulence production activities. By increasing the height of C-D riblets, there is a significant rise in the population of both the prograde and retrograde spanwise vortices. Throughout the diverging region, except for minor modifications in the near-wall area, the population density of both prograde and retrograde spanwise vortices remains relatively unchanged across a significant portion of the boundary layer. Moreover, it exhibits significantly less sensitivity to an increase in the riblet height.

\subsection{Influence of wavelength}
\begin{figure*}[h!]
\includegraphics[width=0.6\textwidth]{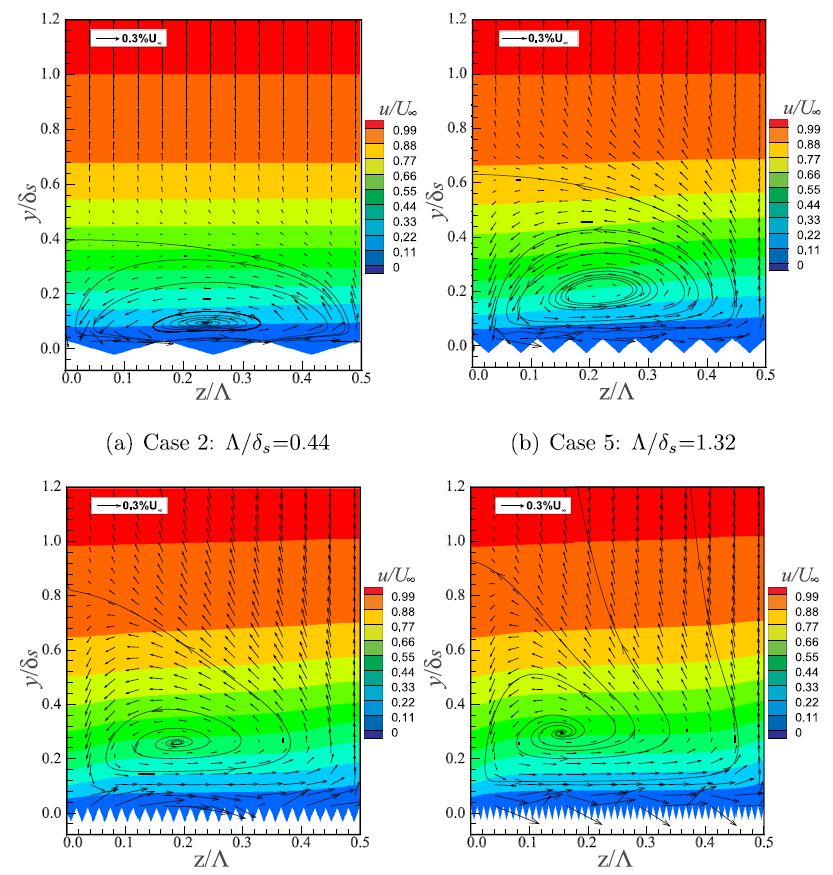}
\caption{The streamwise velocity contours and the vector of in-plane velocity within the cross-stream plane \cite{guo2020secondary}} \label{Fig12}
\end{figure*}

The riblet wavelength is recognized as another important parameter when investigating the physical characteristics of C-D riblets. Guo {\it et al.} examined the impact of varying riblet wavelengths on the laminar boundary layer, with different values of $\Lambda$/$\delta_s$ containing 0.44, 1.32, 3.97, and 6.61 ($\Lambda$ and $\delta_s$ are wavelength and the local boundary layer thickness respectively). They established $\delta$ as the boundary layer thickness using the wall-normal coordinate y. Their report indicates that when the wavelength is small, both the spanwise velocity and the intensity of the secondary flow exhibit relatively low levels of strength (See Fig.~\ref{Fig12}). In situations involving a short wavelength, the restricted spanwise distance for the development of the spanwise velocity between the two planes could cause the requirement for zero spanwise velocity to influence the entire wavelength. As a result, both the spanwise velocity and the intensity of the secondary flow appear relatively weak. 

Xu {\it et al.} employed PIV measurement to examine how riblet wavelength influences the flow field characteristics \cite{xu2018vortical}. The study revealed a notable increase in the amplitude of the induced streamwise velocity across the converging and diverging regions as the riblet wavelength increases. Regardless of the riblet wavelengths, the region exhibiting upwash centred at the converging line is considerably broader than the region with downwash straddling along the diverging line. The influence of riblets on the magnitude of vorticity over the converging region is negligible. Conversely, a larger wavelength induces an increased magnitude of vorticity. Furthermore, it has been observed that variations in riblet wavelength have minimal impact on the spanwise width of vorticity peaks along both the converging and diverging lines. 

Apart from a regular sub-sonic flow, the C-D riblets wavelength variation is also effective in supersonic flow environment. Guo {\it et al.} used direct numerical simulation to investigate the impact of convergent-divergent riblets on the secondary flow induced in supersonic turbulent boundary layers over a 24 degree compression ramp \cite{guo2022direct, guo2022investigation}. They studied the effect of C-D riblets on the secondary rolling motion, momentum transfer, turbulent fluctuations and flow separation with two riblet cases using wavelengths $\Lambda$ of 1.1$\delta$ and 1.65$\delta$ (where $\delta$ is the boundary layer thickness). As the boundary layer advances over the riblet section in the streamwise direction, the magnitude and strength of the secondary rolling motion increase rapidly at first and then seem to rise gradually. The case with $\Lambda$/$\delta$=1.1 exhibited a single roll mode with a size of a half of wavelength, whereas $\Lambda$/$\delta$=1.65 produced a pair of co-rotating vortical structures. Both patterns produce an evident spanwise change in the flow field. According to their findings, the secondary flow increases the average momentum flux and turbulent fluctuations. They acquired that both riblet cases mitigate flow separation with varying vortical structures. Compared to the case without C-D riblets, the area of the separation zone is reduced by 56\% and 38\% for the riblet cases with $\Lambda$/$\delta$=1.1 and $\Lambda$/$\delta$=1.65, respectively. 

\subsection{Fetch length and reversion from rough to smooth surface}
\begin{figure*}[h!]
\includegraphics[width=0.6\textwidth]{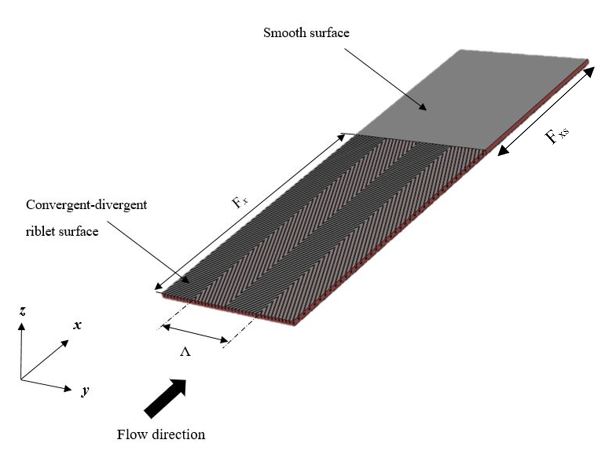}
\caption{Schematic of fetch length and surface reversion from C-D riblets to smooth.} \label{Fig13}
\end{figure*}

\begin{figure*}[h!]
\includegraphics[width=0.6\textwidth]{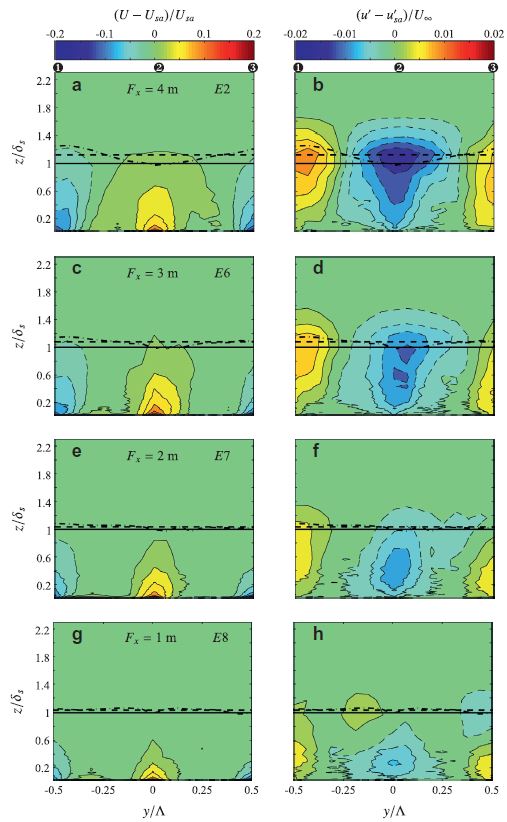}
\caption{Spanwise variation of the mean velocity (left-side images) and turbulence root-mean-squared fluctuations (right-side images) about the spanwise averaged value for convergent-divergent riblets with (a and b) $F_x$ =4m, (c and d) $F_x$ =3m, (e and f) $F_x$ =2m, and (g and h) $F_x$=1m \cite{nugroho2013large}.}
\label{Fig14}
\end{figure*}

The final parameters that we will discuss are the streamwise fetch length ($F_x$) and the streamwise reversion length ($F_{xs}$). The streamwise length fetch ($F_x$) is defined as the rough surface along the streamwise direction where the boundary layer has developed (See Fig.~\ref{Fig13}). Nugroho {\it et al.} \cite{nugroho2013large} evaluated the influence of fetch on the mean velocity and turbulent intensity, by measurements at a fixed streamwise location above C-D riblets and then systematically replacing the upstream rough tiles with a smooth surface. They found that a decrease in the streamwise fetch reduces the extent of regions exhibiting spanwise variation in the mean velocity and turbulent intensity (See Figure~\ref{Fig14}). The induced spanwise variations appear to be more confined closer to the wall for shorter fetch distances. Consequently, the shorter fetch cases show significantly reduced impact on the outer portion of the layer and exhibit notably less spanwise variation in boundary layer thickness. Nevertheless, even at the lowest fetch, the roughness can still impose a discernible three-dimensionality onto the flow.

\begin{figure*}
\includegraphics[width=0.6\textwidth]{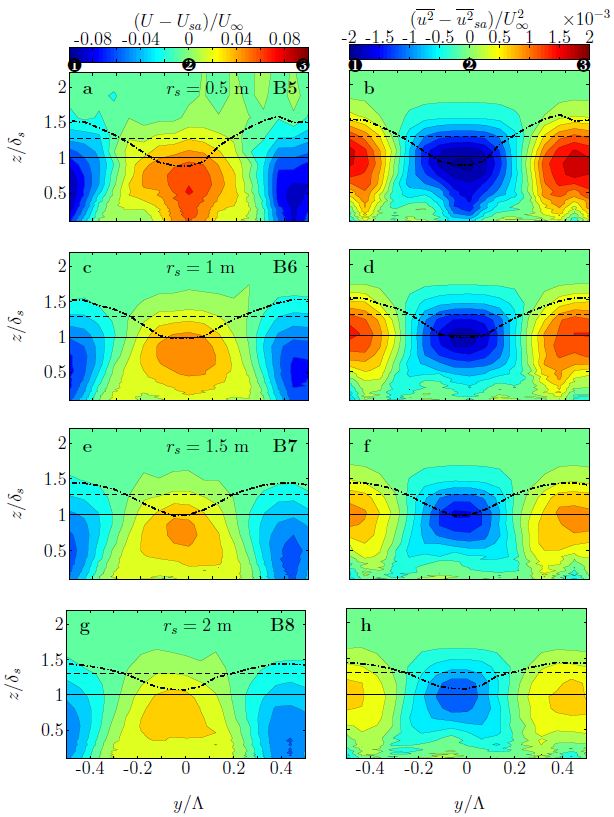}
\caption{Spanwise variation of the mean velocity (left-side images) and turbulence root-mean-squared fluctuations (right-side images) about the spanwise averaged value for the cases with C-D riblets for case B5 $F_{xs}$ =0.5 m (a and b), case B6 $F_{xs}$ =1 m (c and d), case B7 $F_{xs}$ =1.5 m (e and f), case B8 $F_{xs}$=2m (g and h) \cite{nugroho2015highly}.} \label{Fig15}
\end{figure*}

The final component of the discussed parametric study is the effect of a reversion from C-D riblets to a smooth surface. Here streamwise reversion length ($F_{xs}$) is defined as the distance downstream of a step change in surface from C-D riblets to smooth. A general schematic of the reversion from rough surface to smooth is illustrated in Figure~\ref{Fig13}. The findings of Nugroho {\it et al.} indicated that the strength or magnitude of the perturbation decreases as the flow transitions from the rough to the smooth surface (See Figure~\ref{Fig15}). Nevertheless, the large-scale roll-modes persist and a significant spanwise variation is still induced in the boundary layer thickness. Additionally, the size of the roll modes which is related to the size of the spanwise variation region remains unchanged.

\section{The performance of C-D riblets in drag reduction}

\begin{figure*}
\includegraphics[width=0.7\textwidth]{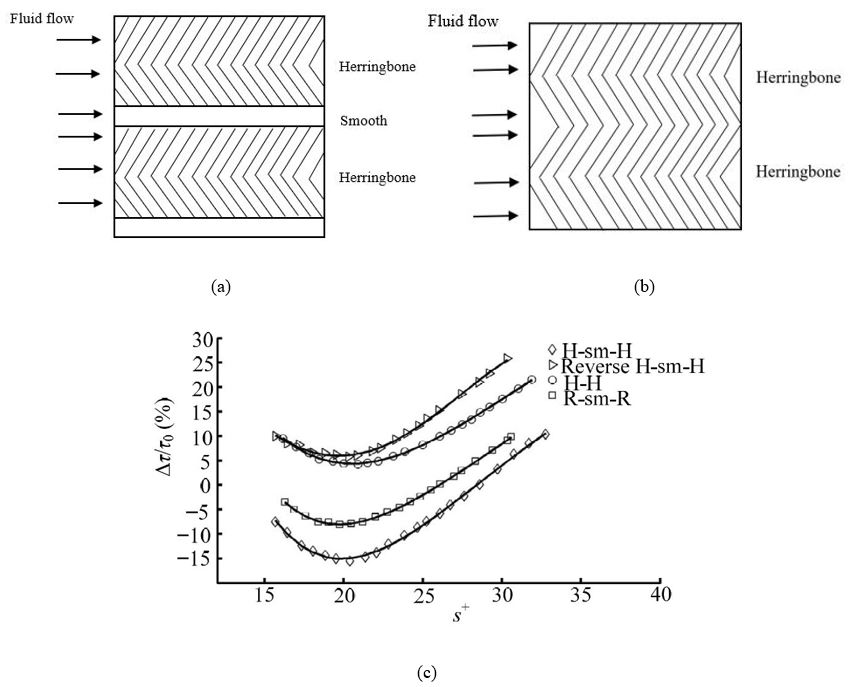}
\caption{Schematic of biomimetic herringbone riblets \cite{chen2013biomimetic}. (a) herringbone-smooth-herringbone and (b) herringbone-herringbone. (c) shows the amount of drag reduction percentage for various riblets arrangements at different viscous scaled space distances.} \label{Fig16}
\end{figure*}

The ability of C-D riblets to modify the mean velocity and turbulent intensity within the turbulent boundary layer has raised questions about its ability to reduce skin friction drag. To address this, Chen at al. used C-D riblets in a pipe to evaluate the feasibility of flow control and drag reduction \cite{chen2013biomimetic, chen2014flow}. Four distinct riblet arrangements, namely herringbone-smooth-herringbone (H-sm-H), reverse herringbone-smooth-herringbone (Reverse H-sm-H), herringbone-herringbone (H-H) and U/V riblets (R-sm-R), were tested (see Fig.~\ref{Fig16}a and \ref{Fig16}b). H-sm-H is an arrangement of a smooth surface gap between two diverging pattern. Reverse H-sm-H combines a smooth surface gap between two converging surface. H-H is named for a C-D surface pattern. Finally, R-sm-R is the traditional microgrooves in which the riblets and valleys make U or V shapes. The fluid velocity for all tested cases ranged from 3 to 8 m/s and the corresponding drag reduction rates are depicted in Fig.~\ref{Fig16}c. The outcomes revealed that the H-sm-H roughness configuration yielded a drag reduction rate of nearly 16\%, which is significantly higher than that of the traditional U/V riblets (R-sm-R) at 6\%. The analysis of the findings demonstrates that the most effective drag reduction by herringbone microgrooves is achieved by the H-sm-H pattern with a diverging arrangement.

Benschop {\it et al.} extended the experiment of Chen {\it et al.} and conducted a direct numerical simulation of a similar C-D riblet pattern inside a turbulent channel flow \cite{benschop2017drag}. They found that the herringbone pattern can either increase or decrease drag depending on the wavelength of the spanwise texture. Specifically, the drag was observed to increase to a maximum of 73\% for narrow feathers with smaller wavelengths. This augmentation is attributed to the C-D riblets, which generate a fluctuating secondary flow comprising two opposite-rotating vortices positioned above the convergent/divergent regions of riblets on average. This robust secondary flow increases the mean and turbulent advective transport, eventually boosting the drag considerable. Wide feathers were observed to exhibit a 2\% drag reduction. Because of their larger width, the C-D riblets generate a secondary flow that affects only a fraction of the overall texture. Thus, their role in augmenting drag is minimal. Most directional riblet texture behaves like a conventional parallel-riblet texture during yaw, with the reduction in turbulent advective transport primarily responsible for the marginal drag reduction achieved. The discrepancies in the turbulent flow drag reduction results between Chen at al. \cite{chen2013biomimetic, chen2014flow} and Benschop {\it et al.} \cite{benschop2017drag}, have raised further questions about the drag reduction capability of the C-D riblets. 

Guo {\it et al.} also used direct numerical simulation to investigate the drag in a turbulent channel flow over C-D riblets \cite{guo2022DNS}. Their results indicate that all riblet cases experienced an augmentation in drag. A closer look at the spanwise drag variation shows that a minor reduction in drag occurs exclusively over a narrow region near the converging line. The rise in total drag results from a significant increase in drag over the diverging section, primarily caused by the downward flow, along with a minor increase in drag between the diverging and converging lines. They decomposed the drag into Reynolds stress and dispersive stress. The major drag was related to Reynolds stress. Subsequently, Guo {\it et al.}, extended their work by studying the laminar channel flow with C-D riblets \cite{guo2022drag}. They examined how factors such as Reynolds number, riblet wavelength and riblet cross-sectional shape influenced drag characteristics. The behavior of drag was studied at different Re numbers (12.5-800). Additionally, their studies showed that the drag increment normalized by the baseline cases of drag coefficient remains relatively stable until Re=100. Afterward, a sudden rise caused by the dispersive velocity arising from the secondary flow. This trend differs significantly from that observed in a laminar channel flow developing over homogeneous roughness, where the normalized drag increment remains consistent up to a much higher Reynolds number due to the lack of secondary flow. However, despite the studies conducted about the effect of C-D riblets on drag reduction, the available information in this field is limited and the detail about the capability of C-D riblets in terms of drag reduction is not yet fully clear; therefore, it is not easy to draw conclusions about the performance of directional riblet texture in drag reduction. Thus, more studies on the effect of C-D riblets on reducing drag at various canonical cases (turbulent boundary layer, turbulent channel flow, and turbulent pipe flow) at matching Reynolds number is worthwhile to give researchers a complete understanding.

\section{Manufacturing techniques for riblets: applications in fundamental research and commercial industries}

In the last few decades there have been many efforts to replicate the nature-inspired riblets, including C-D riblets. The design and manufacture of these riblets have always been considered a challenge due to their small size, varying cross-sectional area and arrangements. The main objective of this section of the paper is to elaborate on the different techniques employed in making riblets. Although most of the discussed techniques are for standard riblets making, they  can be employed to directional riblets. Note that in this section we will discuss both the manufacturing and replication process for both research and development environment and real-world application. 

\subsection{Manufacturing by machining techniques}

The first and most basic manufacturing method is via small-scale machining such as mill or lathe (both manual or automatic via CNC) and directly applied onto fluid flow apparatus (wind tunnel, water tunnel, channel flow, etc) or used as a master model that is replicated via other methods. This technique was used by the pioneer of this field at NASA in the early 1980’s by Walsh et al \cite{walsh1984optimization, walsh1990effect, walsh1982turbulent, walsh1983riblets}. They investigated looking at different riblet cross sections that are cut via different end mill heads (triangular, scallop, etc). Their machining technique is followed by others such as Bechert et al \cite{bechert1997experiments}, Chen et al \cite{chen2013biomimetic}, Nugroho et al \cite{nugroho2013large, nugroho2014roll}, etc.

Apart from the traditional machining process, a more sophisticated machining concept in the form of laser machining is also gaining momentum \cite{kietzig2014laser, basset2022effect, west2018material}. Such a process allows the manufacturing of a surface with accuracy down to the low micro- to high nanoscale and little or no post-processing \cite{west2018material}. However, the challenge with laser machining is the introduction of random and unwanted structures.

\subsection{Manufacturing by Rolling}

\begin{figure*}[h!]
\includegraphics[width=0.65\textwidth]{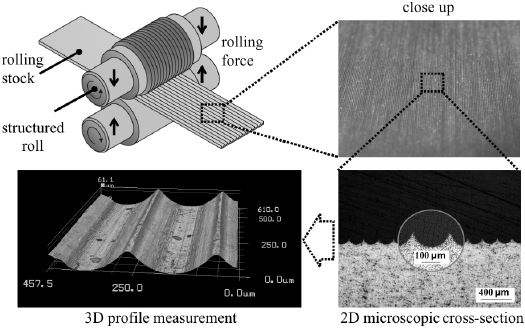}
\caption{Sketch of riblet production by rolling process  \cite{romans2010rolling}.} \label{Fig17}
\end{figure*}

Rolling methods have potential benefits in various fields, including aerospace and fluid transport. The surfaces characterized by micro-scale riblets can be produced by rolling methods \cite{romans2010rolling, klocke2007development, gao2021research}. As seen in Figure~\ref{Fig17}. in general, the process of using two rollers involves rolls extending over a set distance. The material is fed between these rolls. While the rolls rotate, frictional and compressive forces compel the rolling stock to enter the roll gap, resulting in a reduction in its thickness. Within the roll gap, there is a potential opportunity to texture the surface of the rolling stock if the rolls can be equipped with a negative imprint. While rolling the imprint can be effectively transferred onto the rolling stock, thereby achieving the intended profile. 

\subsection{Manufacturing by 3D printing}

\begin{figure*}[h!]
\includegraphics[width=0.55\textwidth]{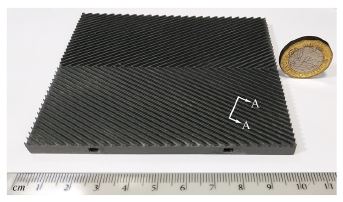}
\caption{ A manufactured model of C-D riblets by 3D printing techniques \cite{xu2019statistical}.} \label{Fig18}
\end{figure*}

In the last 10 years, the use of 3D printing has gained widespread acceptance as a viable method for manufacturing micro-scale devices in research laboratories \cite{balakrishnan20203d}. Various 3D printing-related techniques have been developed to print enclosed microchannels or pipes in many different fields such as medical devices or sensors. Such technology can be utilized to generate riblets that generally have small dimensions. Recent riblet studies by The University of Manchester have shown that such manufacturing techniques are indeed feasible \cite{xu2019statistical, xu2018vortical} and provide an alternative to the more traditional cutting and rolling technique. Their advanced 3D printer is capable of generating accurate riblets due to its high precision level (in the vicinity of 25 microns). The main challenge with this method, however, is the relatively longer time duration needed to generate a plate of riblets. This is because the 3D printer would not just be generating the riblets, it would also need to assemble the thicker base surface. Figure~\ref{Fig18} exemplifies a model of convergent-divergent riblets, manufactured using the 3D printing method.

\subsection{Manufacturing by film}

One of the earliest mass-produced applied riblets were manufactured by the 3M company in the 1980’s in the form of thin plastic film with an adhesive backing. Since then the riblet technology has been used on pipe flow \cite{koury1995drag}, ship model \cite{choi1989tests}, operating aircraft wings/surface \cite{mclean1987flight, szodruch1991viscous}, full-sized ship hulls \cite{letcher1987stars}, and wind turbine \cite{chamorro2013drag}. Such types of riblet films can be manufactured by applying polydimethylsiloxane (PDMS) onto a master cut and can be shaped into a flexible and thin riblet film that can be applied to the surface \cite{han2002micro, lee2008decrement}.

\subsection{Manufacturing by UV paint}

\begin{figure*}[h!]
\includegraphics[width=0.65\textwidth]{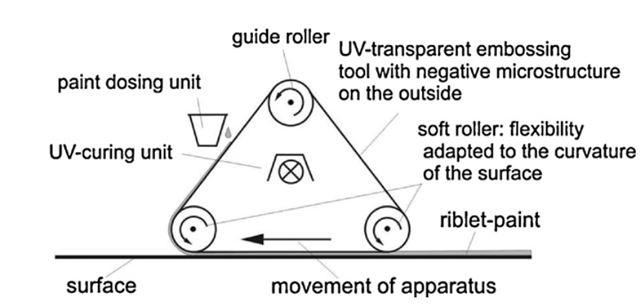}
\caption{Paint application apparatus \cite{stenzel2011drag}} \label{Fig19}
\end{figure*}

Another novel method to apply riblets over a surface with a wide range area is via the paint application technique. This method was first introduced by the Fraunhofer Institute for Manufacturing Technology and Applied Materials Research (IFAM) in Bremen, Germany \cite{stenzel2011drag}. Here they applied a paint application apparatus with a flexible microstructure endless belt that is guided over three rollers (see Figure~\ref{Fig19}). As the apparatus moves over a surface, a unique mix of paint that consists of a two-component polyurethane material is applied and also cured by the built-in UV lamp inside the apparatus. Such a technique has been applied in a laboratory for fluid flow experiments and it can produce good-quality riblets \cite{gruneberger2011drag, chen2014investigation}. 

\section{Can riblets be applied commercially?}

As per the report by Spalart and McLean \cite{spalart2011drag}, the primary challenges associated with the utilization of riblets for flow control or drag reduction in practical engineering contexts like aircraft and ships primarily revolve around cost factors related to production and labor. Furthermore, riblets are prone to erosion due to substantial fluctuations in dirt, pressure, and temperature. Therefore, the present challenges are not primarily due to fluid mechanics but instead pertain to various aspects within the broader domain of engineering, encompassing aspects such as painting methods, materials, and manufacturing processes. However, despite this, recent progress and field studies on real-world applications by Airbus, Microtau, BASF, and others may provide solutions for the issues mentioned above.

\section{Conclusion}

In conclusion, the primary aim of this review paper is to offer an in-depth understanding of the various aspects of innovative bio-inspired convergent-divergent riblets. Unlike longitudinal riblets, convergent-divergent (C-D) patterns uniquely demonstrate the ability to influence boundary layer flows, vortical structures, near-wall motions, and the onset of flow separation. Moreover, they can specifically target the dominant large and very large-scale structures that dictate the boundary layer in high Reynolds number conditions. 

This study delves into the flow patterns generated by C-D riblets. It's crucial to highlight that the vortex generation principles exhibited by convergent-divergent riblets differ from those seen in traditional riblet designs. The configuration of the converging and diverging orientations of the riblets, in contrast to the streamwise flow, results in varied flow patterns. These include helical and rotational motions within the riblet valleys, upward and downward movements, and counter-rotating roll modes. The unique vortex generation mechanisms of C-D riblets significantly influence the boundary layer.

A central focus of this research was on the key physical parameters associated with C-D riblets. These parameters include the yaw angle, wavelength, viscous-scaled riblet height, fetch length, and the transition from C-D riblets to a smooth surface. These factors can profoundly affect the intensity of spanwise variation and the formation of flow structures. 

From the research conducted by various scholars, it's clear that C-D riblets hold potential for effective drag reduction. However, despite the existing studies on the effects of C-D riblets on drag reduction, our understanding in this area remains limited. This limitation makes it difficult to draw definitive conclusions about the efficacy of C-D riblets in reducing drag. Therefore, more research on the drag-reducing effects of C-D riblets would be beneficial, providing a fuller understanding for researchers in this field. 

Given the riblets' small size and concerns about their durability and the potential for dust accumulation in the valleys, the production methods for riblets have always posed challenges for both researchers and the commercial sector. Convergent-divergent riblets are no exception. In this paper, we discuss various manufacturing techniques, including machining, rolling, 3D printing, film, and UV paint methods. This section aims to inform researchers and industries about the latest advancements in C-D riblet production. Additionally, we explored the potential commercial applications of convergent-divergent riblets, with a particular focus on the aviation industry. 

In summary, despite challenges related to their production and practical application in the commercial realm, C-D riblets present a novel passive flow control strategy. Their unique ability to positively impact fluid flow, due to their distinct riblet design, positions them as a promising avenue for future flow control methods.

\begin{acknowledgments}
This work was supported by 'Human Resources Program in Energy Technology' of the Korea Institute of Energy Technology Evaluation and Planning (KETEP), granted financial resource from the Ministry of Trade, Industry \& Energy, Republic of Korea (no. 20214000000140). In addition, this work was supported by the National Research Foundation of Korea (NRF) grant funded by the Korea government (MSIP) (no. 2019R1I1A3A01058576).
\end{acknowledgments}

\section*{Data Availability}
The data that supports the findings of this study are available within this article.


%


\end{document}